# Investigation of the effectiveness of applying ChatGPT in Dialogic Teaching Using Electroencephalography

Jiayue Zhang
*School of Physics and Information Technology*
*Shaanxi Normal University*
Xi'an, China
zhangjiayue@snnu.edu.cn

Yiheng Liu
*School of Physics and Information Technology*
*Shaanxi Normal University*
Xi'an, China
lyh37779816@gmail.com

Wenqi Cai
*School of Psychology*
*Shaanxi Normal University*
Xi'an, China
Wenqi_Cai@snnu.edu.cn

Lanlan Wu
*School of Physics and Information Technology*
*Shaanxi Normal University*
Xi'an, China
2823778952@qq.com

Yali Peng
*School of Computer Science*
*Shaanxi Normal University*
Xi'an, China
pengyl@snnu.edu.cn

Jingjing Yu
*School of Physics and Information Technology*
*Shaanxi Normal University*
Xi'an, China
yujj@snnu.edu.cn

Senqing Qi
*Key Laboratory of Modern Teaching Technology, Ministry of Education*
*Shaanxi Normal University*
Xi'an, China
qisenqing@126.com

Taotao Long*
*Faculty of Artificial Intelligence in Education*
*Central China Normal University*
Wuhan, China
taotaolong@ccnu.edu.cn

Bao Ge*
*Key Laboratory of Modern Teaching Technology, Ministry of Education, School of Physics and Information Technology*
*Shaanxi Normal University*
Xi'an, China
bob_ge@snnu.edu.cn

*Abstract*—In recent years, the rapid development of artificial intelligence technology, especially the emergence of large language models (LLMs) such as ChatGPT, has presented significant prospects for application in the field of education. LLMs possess the capability to interpret knowledge, answer questions, and consider context, thus providing support for dialogic teaching to students. Therefore, an examination of the capacity of LLMs to effectively fulfill instructional roles, thereby facilitating student learning akin to human educators within dialogic teaching scenarios, is an exceptionally valuable research topic. This research recruited 34 undergraduate students as participants, who were randomly divided into two groups. The experimental group engaged in dialogic teaching using ChatGPT, while the control group interacted with human teachers. Both groups learned the histogram equalization unit in the information-related course "Digital Image Processing". The research findings show comparable scores between the two groups on the retention test. However, students who engaged in dialogue with ChatGPT exhibited lower performance on the transfer test. Electroencephalography data revealed that students who interacted with ChatGPT exhibited higher levels of cognitive activity, suggesting that ChatGPT could help students establish a knowledge foundation and stimulate cognitive activity. However, its strengths on promoting students' knowledge application and creativity were insignificant. Based upon the research findings, it is evident that ChatGPT cannot fully excel in fulfilling teaching tasks in the dialogue teaching in information related courses. Combining ChatGPT with traditional human teachers might be a more ideal approach. The synergistic use of both can provide students with more comprehensive learning support, thus contributing to enhancing the quality of teaching.

## I. INTRODUCTION

The emergence of Large Language Models (LLMs), represented by the GPT family [18], has facilitated the development of Natural Language Processing (NLP) [6] applications. ChatGPT is representative of the Large Language Models (LLMs) [5], including GPT-3.5 and GPT-4.

In the field of education, dialogic teaching serves as a prevalent instructional method that underscores the importance of dialogue and interaction between teachers and students [1]. This approach aims to stimulate students' thinking, comprehension, and active participation in the learning process. Within dialogic teaching, mutual communication occurs between educators and students, encompassing the exchange of questions, sharing of viewpoints, explanation of concepts, and the deepening of students' understanding through dialogue. The emergence of ChatGPT [5] has propelled the application of dialogues with artificial intelligence across various domains [6], particularly within the realm of education. ChatGPT is commonly applied in question-and-answer scenarios resembling dialogic teaching, which can remember the history of the conversation, integrate contextual information, and thereby understand and respond. When students engage in a dialogue with ChatGPT, it can better understand the context of new questions based on previous conversation content. Leveraging information and patterns from its training data, ChatGPT can provide relevant answers or explanations, generating appropriate responses [7]. ChatGPT can also simplify and explain complex concepts, making them easier to understand, thereby assisting students in overcoming learning difficulties [8]. Considering the characteristics of ChatGPT research, it becomes evident that ChatGPT closely aligns with the traditional role of a human teacher in dialogic teaching. Therefore, investigating whether ChatGPT can assume the role of a human teacher in dialogic teaching, aiding students in resolving queries and providing support in dialogic instruction, represents a crucial and valuable research topic.

Therefore, this research uses experimental methodologies, complemented by data collection and analysis techniques derived from educational neuroscience, to investigate the effectivene of using ChatGPT in dialogic teaching in information related courses for undergraduates. A total of 34 college students participated in the study, with random assignment to two groups: the experimental group engaging in conversation with ChatGPT, and the control group conversed with human teachers for learning. The chosen topic for the dialogic teaching interactions, conducted between students and teacher/ChatGPT, was the histogram equalization unit in the "Digital Image Processing" course. We analyzed the learning performance of the two groups with regards of the retention and transfer tests scores. Additionally, we compared the differences in EEG patterns during the dialogue process between the two groups. This study aspires to offer meaningful insights into the utilization of ChatGPT in dialogic teaching, providing specific guidance for the future integration of educational technology and artificial intelligence.

## II. LITERATURE REVIEW

Dialogic Teaching, as a common teaching method [1][2], has been extensively studied in the field of education, with previous research primarily focusing on conventional human teachers [3][4]. The previous research on ChatGPT in the field of education mainly focused on answering questions or explaining concepts. Pardos et al. [7] presented ChatGPT with questions from the OpenStax textbook in a question-and-answer format to explore ChatGPT's ability to answer math questions. However, the study was limited to a single question. In the field of physics, Lehnert et al. [8] investigated ChatGPT's ability to explain concepts related to combination, mixture, and synthesis during teaching. However, there was no in-depth exploration of ChatGPT's ability to explain continuously related concepts when prompted. It can be observed that the previous studies on the application of ChatGPT in the field of education were relatively simple. Therefore, we delved deeper into the application of ChatGPT in dialogic teaching, investigating its ability to respond to a series of continuous related questions from students or explain a series of continuous related concepts,

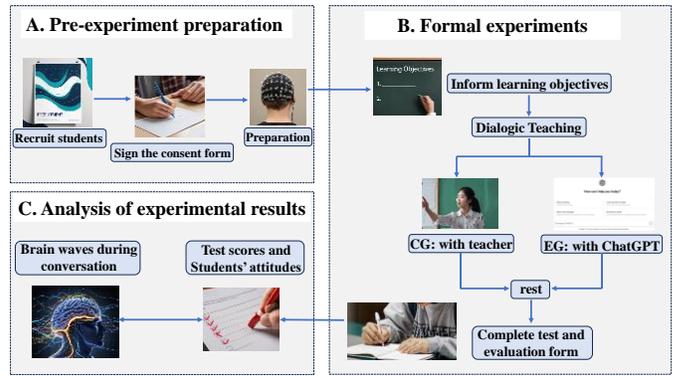

Fig. 1. Experimental Flowchart

and effectively communicate with students to excellently accomplish teaching tasks.

## III. METHODS

This research employed an experimental research design, recruiting and selecting 34 undergraduate students as participants. The independent variable is the teacher/ChatGPT in dialogue-based teaching, while the dependent variables were the learning performance of the two groups of students and the Electroencephalography (EEG) patterns during the dialogue process.

### A. Context and Participants

We recruited 34 undergraduate student participants (18 males, 16 females) from a large research university in Northwest China. Their mean age was 21.56 years (SD = 1.73). All students majored in science and engineering. Up to 30% were sophomore, 30% were junior, and 40% were senior. None of them had previously taken a course in "Digital Image Processing". According to the Edinburgh Handedness Inventory [9], all participants were right-handed and had not been diagnosed with any neurological disorders or learning disabilities. The study strictly adhered to ethical review procedures, and participants' identities were kept completely confidential. Before the study commenced, participants were fully informed and consented to participate. Participants received monetary rewards for their participation in the study.

In this study, participants were randomly assigned to two groups. The control group (CG) engaged in dialogic teaching with a human teacher majored in electronic information. CG consisted of 17 students (9 males, 8 females) with a mean age of 21.47 years (SD = 1.74). The experimental group (EG) engaged in dialogic teaching with ChatGPT and constituted the experimental group. EG consisted of 17 students (9 males, 8 females) with a mean age of 21.65 years (SD = 1.77).

### B. Experimental Design

The experimental procedure is illustrated in Fig. 1, which consists of three main stages. Firstly, there is the pre-experiment preparation stage (Fig. 1. A.), followed by the formal experimental stage (Fig. 1. B.), and finally, the post-experiment result analysis stage (Fig. 1. C.). The following sections will provide detailed descriptions of these three parts.

*1)* Pre-experiment preparation

In the CG who interacted with the human teacher, the role of the teacher was assumed by a trained graduate student. This graduate student is a second-year student majoring in electronic information, possessing a strong academic background and extensive foundational knowledge. He had training on conducting dialogue-based teaching as a teacher. In the EG who interacted with ChatGPT, there was an experimental assistant assigned to address technical issues. However, the assistant did not intervene or answer any questions related to the learning content. Prior to the formal experiment, the experimental assistant provided instructions to the EG students on how to use ChatGPT.

The pre-experiment preparation stage is shown in Fig. 1. A. The undergraduate student participants are majoring in science and engineering related disciplines and have not taken the course "Digital Image Processing". Before the experiment began, the teacher/assistant introduced the experimental content to the participants, and the participants signed informed consent forms. Next, preparations for the formal experiment were made. The preparations before the formal experiment are as follows: Students completed pre-experiment test questions focusing on knowledge related to histogram equalization within the "Digital Image Processing" course, and watching relevant videos and studying materials to prepare for the course. Then the students wore the EEG caps for the formal experimental stage. Before the dialogue begins for both groups of students, the teacher/assistant informs the students that the formal experimental session for dialogic teaching will last between three to ten minutes.

### 2) Formal experiments

The formal experimental stage is depicted in Fig. 1. B. After the formal experiment began, the teacher/assistant informed the students of the learning objectives for this class session. For the students in EG, before they started the dialogue with ChatGPT, the experimental assistant informed ChatGPT: "Assume you are a lecturer teaching undergraduate's digital image processing course. Your assistance is required to facilitate students in learning the unit on histogram equalization " Subsequently, during the formal experimental phase of dialogic teaching, Both groups of students asked the teacher/ChatGPT all the questions they did not understand, such as: "What is the purpose of equalization? Why is normalization performed?" The teacher/ChatGPT answered all the questions asked by the students. Throughout the entire dialogue process for both groups of students, the EEG equipment continuously recorded the students' brainwaves in real time. After the dialogue-based teaching, students took a break for three to five minutes, followed by completing the test questions and filling out evaluation forms.

### 3) Analysis of experimental results

The evaluation and analysis of the experimental results are illustrated in Fig. 1. C. The first part involves evaluating student learning outcomes, including objective comparisons of pre-test and post-test scores, as well as subjective evaluations from both groups of students. Following the guidelines of educational psychology [10], two tests were used to assess learning outcomes: a knowledge retention test and a knowledge transfer test. The test questions were modified versions of standardized tests developed by several experts in the field. The knowledge retention test consists of 2 true/false questions and 9 multiple-choice questions, totaling 40 points. It is used to assess students' memory of the learned material, thereby measuring the basic level of knowledge acquisition. The knowledge transfer test consists of 1 calculation question, totaling 40 points. Students need to apply the learned knowledge to new contexts in order to assess their deeper level of understanding. We employed a 5-point Likert scale [11] designed assessment form. This form is used to assess the extent to which the teacher and ChatGPT were able to address student questions, as well as the degree of satisfaction of both groups of students with the dialogue with the teacher and ChatGPT. The second part involves the analysis and study of EEG waves in the theta frequency band during the dialogue process of both groups of students. The specific steps of EEG recording and processing analysis are detailed in III. C.

### C. EEG measurement and processing during conversation

This study utilized the 10-20 system introduced by the International Federation of Clinical Neurophysiology [12]. Ag/AgCl electrodes were placed at 64 scalp locations to record raw EEG data, with the left mastoid process serving as the reference point and an electrode for grounding on the forehead. The EEG signals were amplified using the SynAmps2 amplifier (Neuroscan, Herndon, VA, USA), with all electrode impedances maintained below 5 kΩ. The sampling rate was set to 1 kHz. Data preprocessing was conducted using the EEGLAB toolbox [13] in MATLAB. The data were downsampled to 512 Hz, followed by a global average reference. Filtering of the data was performed within the range of 0.5 to 60 Hz. Using independent component analysis (ICA) technique to enhance the quality and accuracy of signals. To maintain consistency in EEG data analysis, the dialogue processes of all students were truncated, and a fixed length of 200 seconds was selected as the basis data for the subsequent analysis.

The study utilizes Power Spectral Density (PSD) as the method for extracting frequency domain features from EEG signals [14]. Welch's method is employed to extract Power Spectral Density. This method initially divides the signal of length N into segments and applies windowing techniques to obtain. Subsequently, the discrete Fourier transform (FFT) is applied to each windowed segment to calculate the power spectral density estimate for each segment. Finally, the average power spectral density estimate across all segments is computed to obtain the final power spectral density estimate. Based on previous research results, the frequency range selected for the Theta band is 4-8Hz [15].

## IV. RESULTS

### A. The assessment of learning outcomes

We conducted comparative analyses and evaluations of the effectiveness of the two groups in dialogue-based teaching, including objective score analysis and subjective evaluations from students. Objective score analysis involved comparing scores between pre-tests and post-tests, as well as comparing performance on retention test and transfer test items between the two groups. Subjective evaluations from students included assessments of their perception of problem-solving abilities and level of enjoyment.

### 1) Learning achievement

TABLE I. MEAN SCORES, STANDARD DEVIATIONS (SD), AND T-TEST STATISTICS FOR PRETEST SCORES OF TWO GROUPS OF STUDENTS

|  | CG Mean (SD) | EG Mean (SD) | t | p |
|---|---|---|---|---|
| Pre-test | 20.06 (6.26) | 20.76 (6.86) | -0.313 | 0.756 |

TABLE II. MEAN SCORES, STANDARD DEVIATIONS (SD), AND T-TEST STATISTICS FOR POSTTEST KNOWLEDGE RETENTION AND TRANSFER TEST SCORES OF TWO GROUPS OF STUDENTS

| Post-test | CG Mean (SD) | EG Mean (SD) | t | p |
|---|---|---|---|---|
| Retention test | 32.53 (4.87) | 32.29 (5.42) | 0.133 | 0.895 |
| Transfer test | 33.06 (7.60) | 28.59 (8.54) | 1.612 | 0.117 |

Before the experiment, we conducted tests on the two groups of students (total score 40 points) to examine whether there were significant differences in prior knowledge between the two groups. We collected and calculated the average scores, standard deviations, and t-test statistics for the two groups of students, and the results are shown in TABLE I. From the pre-test scores, the mean score (SD) for CG is 20.06 (6.26), and for EG is 20.76 (6.86). Considering the results of the t-test, the difference in pre-test scores between the two groups of students is not significant, the scores are very close, indicating that the prior knowledge of the two groups of students is basically consistent.

After the experiment, knowledge retention and transfer tests were conducted on both groups of students (with a total score of 40 points each). Likewise, we collected and calculated the average scores, standard deviations (SD), and t-test statistics for each group of students, and the results are shown in TABLE II. We also plotted box plots of scores on knowledge retention and transfer test Scores, as shown in Fig. 2. The retention test results indicate that the mean score (SD) for CG is 32.53 (4.87), reflecting an increase of 12.47 points compared to the pre-test score. For EG, the mean score (SD) is 32.29 (5.42), which is an increase of 11.77 points compared to the pre-test score. Students' post-test retention scores show a high increase compared to the pre-test scores for both groups. Considering the t-test results, the average scores of the two groups of students are very close, and the difference is not significant. The transfer test results indicate that the mean score (SD) for CG is 33.06 (7.60), which is an increase of 13 points compared to the pre-test score. For EG, the mean score (SD) is 28.59 (8.54), indicating an increase of 7.83 points compared to the pre-test score. Both groups of students show an increase in post-test transfer scores compared to the pre-test scores, but the increase in CG is significantly higher than that in EG. The average score of CG is higher than that of EG, but the t-test results do not show a significant difference.

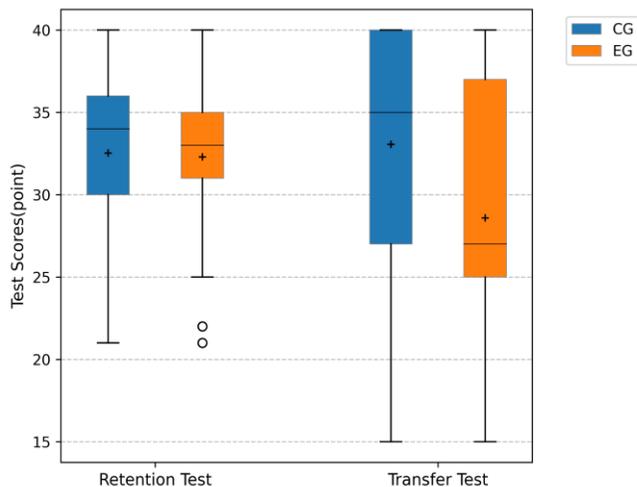

Fig. 2. Box plots of Posttest Knowledge Retention and Transfer Test Scores of Two Groups of Students

TABLE III. THE EVALUATION OF PROBLEM-SOLVING PROFICIENCY BY THE TWO GROUPS OF STUDENTS IS PRESENTED IN THE TABLE, INDICATING THE NUMBER OF STUDENTS FOR EACH LEVEL OF RATING. A RATING OF 5 DENOTES COMPLETE RESOLUTION, WHILE 1 SIGNIFIES UNRESOLVED ISSUES. SCORES FROM 5 TO 1 REPRESENT THE DEGREE OF PROBLEM RESOLUTION, RANGING FROM THOROUGH RESOLUTION TO UNRESOLVED

|  | 5 | 4 | 3 | 2 | 1 |
|---|---|---|---|---|---|
| CG | 5 | 11 | 1 | 0 | 0 |
| EG | 1 | 13 | 2 | 1 | 0 |

TABLE IV. THE TABLE LISTS TWO GROUPS OF STUDENTS' EVALUATIONS OF THEIR LIKING FOR THE DIALOGUE, INDICATING THE NUMBER OF STUDENTS FOR EACH RATING LEVEL. A SCORE OF 5 INDICATES VERY MUCH LIKED, WHILE A SCORE OF 1 INDICATES VERY MUCH DISLIKED. SCORES RANGE FROM 5 TO 1, REPRESENTING VARYING DEGREES OF LIKING, FROM VERY MUCH LIKED TO VERY MUCH DISLIKED

|  | 5 | 4 | 3 | 2 | 1 |
|---|---|---|---|---|---|
| CG | 11 | 6 | 0 | 0 | 0 |
| EG | 2 | 8 | 6 | 1 | 0 |

*2) Students' attitudes*

We conducted subjective evaluations from both groups of students regarding the dialogic teaching approach, focusing on their perceived problem-solving proficiency and preference. TABLE III. and TABLE IV. present the evaluations of these two indicators by the two groups of students. The results of the demographic analysis from both tables indicate that, in terms of problem-solving proficiency and preference, students from CG all provided higher evaluations. Teacher demonstrates a higher level of proficiency in solving students' queries, and students prefer learning through dialogue with teachers.

*B. Analysis of EEG data during conversation*

We collected and calculated the mean, standard deviation (SD), and t-test statistics of the spectral power in the Theta and Gamma frequency band at electrode points F3, F4, Fz, C3, C4, Cz, P3, P4, and Poz during student dialogue. The results are shown in TABLE V. and TABLE VI. , and the box plot of the spectral power in the Theta and Gamma frequency band is depicted in Fig. 3. and Fig. 4.

*1) EEG Theta Activity Results*

From TABLE V. and Fig. 3., it can be observed that the spectral power in the Theta frequency band of EG students is higher than that of CG students. Considering the t-test statistical results, except for the Cz ($p = 0.091 > 0.05$) and Poz ($p = 0.064 > 0.05$) channels, the t-test results for all other channels show $p < 0.05$, with the C4 channel showing $p < 0.01$. This indicates that there is a significant difference in the theta-band spectral power between the two groups of students during the dialogue process, with students in EG showing more active theta-band EEG activity.

*2) EEG Gamma Activity Results*

From TABLE VI. and Fig. 4., it can be observed that the students conversing with the teacher and with ChatGPT presented some differences in the different electrode channels. For the F3 (13.57 > 8.06), F4 (11.00 > 7.49), C3 (10.52 > 9.38) and C4 (4.05 > 3.89) channels, students conversing with the teacher showed stronger Gamma band activity, and for the rest of the channels, the spectral power of the Gamma band of the brainwaves was higher in the students conversing with ChatGPT than in the students conversing with the teacher. From the statistical results of t-test with p-value, there is no more significant difference between F3, F4, Fz, C3, C4, Cz and P3 channels, and there is a difference between P4 and Poz channels ($p < 0.05$), and the students in EG group clearly show higher Gamma band activity, which is more significant difference compared to the students who dialogued with teachers.

TABLE V. MEAN, STANDARD DEVIATION (SD), AND T-TEST STATISTICS OF THETA BAND SPECTRAL POWER FOR EACH ELECTRODE POINT IN THE TWO STUDENT GROUPS

|     | CG Mean (SD) | EG Mean (SD) | t | p |
| --- | --- | --- | --- | --- |
| F3  | 3.42（1.88） | 5.28（2.72） | -2.277 | 0.030 |
| F4  | 3.11（1.35） | 3.99（2.00） | -1.486 | 0.148 |
| Fz  | 3.89（1.64） | 5.11（2.51） | -1.644 | 0.111 |
| C3  | 1.78（1.20） | 2.54（1.42） | -1.640 | 0.111 |
| C4  | 1.39（0.49） | 2.22（0.97） | -3.132 | 0.004 |
| Cz  | 1.92（0.73） | 2.71（1.70） | -1.745 | 0.091 |
| P3  | 2.01（1.08） | 2.30（1.23） | -0.712 | 0.482 |
| P4  | 2.43（1.44） | 4.27（2.42） | -2.646 | 0.013 |
| POz | 2.94（1.61） | 4.96（3.98） | -1.920 | 0.064 |

TABLE VI. MEAN, STANDARD DEVIATION (SD), AND T-TEST STATISTICS OF GAMMA BAND SPECTRAL POWER FOR EACH ELECTRODE POINT IN THE TWO STUDENT GROUPS

|     | CG Mean (SD) | EG Mean (SD) | t | p |
| --- | --- | --- | --- | --- |
| F3  | 13.57（15.76） | 8.06（5.61） | 1.149 | 0.262 |
| F4  | 11.00（11.80） | 7.49（6.80） | 0.907 | 0.373 |
| Fz  | 15.11（16.78） | 23.37（18.98） | -1.178 | 0.250 |
| C3  | 10.52（10.21） | 9.38（13.23） | 0.249 | 0.806 |
| C4  | 4.05（3.49） | 3.89（3.70） | 0.115 | 0.909 |
| Cz  | 2.44（1.70） | 4.54（6.58） | -1.150 | 0.261 |
| P3  | 8.62（8.08） | 11.53（10.26） | -0.809 | 0.427 |
| P4  | 4.54（3.27） | 13.73（16.08） | -2.094 | 0.047 |
| POz | 8.78（7.36） | 24.92（20.95） | -2.701 | 0.012 |

## V. DISCUSSION

From the test results, ChatGPT performed significantly better on helping students acquire basic knowledge concepts than human teacher, but not in promoting knowledge transfer, such as knowledge application and innovation. Additionally, students believed that human teachers could better guide their problem-solving and preferred to having dialogues with human teachers. Previous studies have shown that the activity of Theta band in the frequency domain of the brain increases in the frontal region when the working memory load increases [16]. Osipova et al [17] emphasized the association between theta activity and the optimal neuronal dynamics of synaptic plasticity, this suggests that theta activity promotes memory encoding activity, i.e., students storing learned knowledge and successfully using it in later memory retrieval. Since students showed higher theta-frequency domain activity during dialogue with ChatGPT, this may reflect the active involvement of EG students' brains in the learning and memory encoding process, helping to strengthen the storage and retrieval of relevant memories.

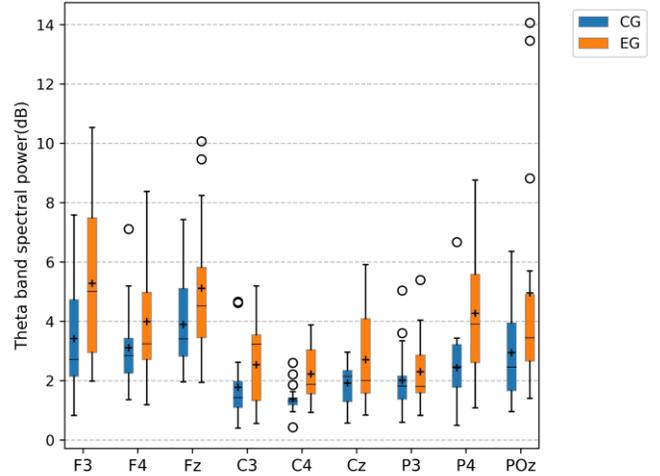

Fig. 3. Boxplots of Theta Band Spectral Power for Each Electrode Point for Two Groups of Students

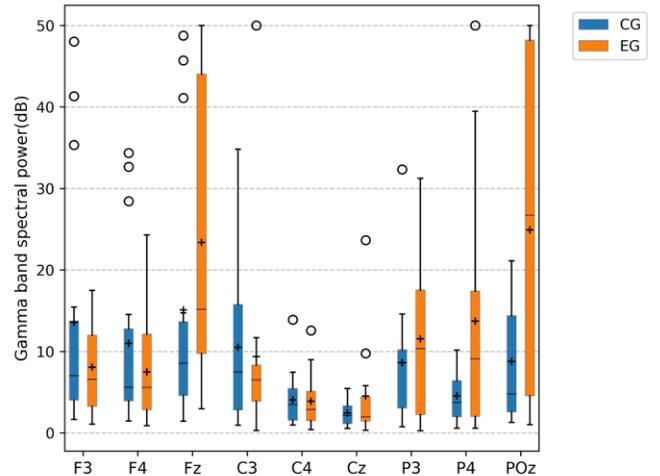

Fig. 4. Boxplots of Gamma Band Spectral Power for Each Electrode Point for Two Groups of Students

It is known from a large number of previous studies [19][20][21][22][23] that strong sustained Gamma band activity is associated with feature binding and selective attention. Several studies [23][24][25] have shown that Gamma band activity reflects an increased level of attention, i.e., an increase in Gamma power reflects an increase in attention, which facilitates memory encoding and retrieval. Osipova et al. [17] identified Gamma activity during a memory task, and an increase in Gamma band power during memory encoding and retrieval likely reflected an increase in neuronal synchronization. The results of the present study showed that in the F3, F4, C3, and C4 channels, students conversing with the teacher showed stronger Gamma band activity, and in the rest of the channels, students conversing with ChatGPT demonstrated higher Gamma band activity, especially in the P4 and Poz channels the students in the EG group were significantly higher than the students in the CG group, and the difference was more significant. Overall, the EG group as a whole showed higher Gamma band activity, and this activity indicates an increase in the attention span of the students in the EG group.

Overall, ChatGPT excelled in assisting students in acquiring basic knowledge concepts, with students showing more focused attention and active thinking during dialogue with ChatGPT. This is not only because interactive teaching has the advantage of promoting student thinking and engagement in the learning process [2][3][4] but also perhaps because ChatGPT can provide personalized real-time learning support, encouraging students to ask questions, discuss viewpoints, thus promoting multi-perspective thinking and stimulating students' thinking activity. However, ChatGPT failed to provide sufficient support when knowledge application and innovation are required, reflecting its limitations. ChatGPT lacks proficiency in knowledge application and innovation, as well as emotional intelligence, unable to understand students' emotional states like human teachers.

This study yields several limitations. First, the course selected for this study is a specialized course in university science and engineering, with the participants primarily being college students. That might threaten the generalization of the findings. Second, only one question was used to assess the knowledge transfer test, more objective measurements on students' learning performance are required. In future research, more extensive and diversified exploration can be conducted, such as covering other disciplinary fields, expanding the scope of participants, further expanding assessment methods for learning outcomes, and more comprehensively evaluating the potential value of ChatGPT in the field of education.

## VI. CONCLUSION

This study provided substantial insights into the application of ChatGPT in teaching. Whether ChatGPT can excel in teaching tasks as brilliantly as traditional human teachers, or even surpass them, remains a matter of debate.

## ACKNOWLEDGMENT

This research was supported by the National Natural Science Foundation of China (No. 61976131, 62307017) and the Shaanxi Province Department of Science and Technology under Grant 2024SF-YBXM-064.